%% file: main.tex
\pdfoutput=1

\documentclass[11pt]{article}

\usepackage[preprint]{acl}

\usepackage{times}
\usepackage{latexsym}
\usepackage{float}
\usepackage{booktabs}
\usepackage{multirow}

\usepackage[T1]{fontenc}

\usepackage[utf8]{inputenc}

\usepackage{microtype}

\usepackage{inconsolata}

\usepackage{graphicx}
\usepackage{tikz}

\usepackage{amsmath}

\usepackage[most]{tcolorbox}

\title{Zero-Shot Dense Retrieval with Embeddings from Relevance Feedback}

\author{
 \textbf{Nour Jedidi\textsuperscript{1}} \quad
 \textbf{Yung-Sung Chuang\textsuperscript{2}} \quad
 \textbf{Leslie Shing\textsuperscript{1}} \quad
 \textbf{James Glass\textsuperscript{2}} \quad 
\\
 \textsuperscript{1}MIT Lincoln Laboratory \quad
 \textsuperscript{2}Massachusetts Institute of Technology
\\
\texttt{\{nour.jedidi, leslie.shing\}@ll.mit.edu} \quad
\texttt{\{yungsung, glass\}@mit.edu}
}

\begin{document}
\maketitle

\begin{abstract}
\input{sections/0_abstract}
\end{abstract}
\input{sections/1_intro}

\input{sections/2_method}
\input{sections/3_experiments}
\input{sections/4_ablation}
\input{sections/5_distill_rede}
\input{sections/6_rede_v_passage_reranking}
\input{sections/7_related_work}
\input{sections/8_conclusion}

\input{sections/limitations}
\input{sections/awknowledgements}

\bibliography{main}

\newpage
\appendix
\input{sections/appendix}

\end{document}

%% file: sections/0_abstract.tex
Building effective dense retrieval systems remains difficult when relevance supervision is not available. Recent work has looked to overcome this challenge by using a Large Language Model (LLM) to generate hypothetical documents that can be used to find the closest real document. However, this approach relies solely on the LLM to have domain-specific knowledge relevant to the query, which may not be practical. Furthermore, generating hypothetical documents can be inefficient as it requires the LLM to generate a large number of tokens for each query. To address these challenges, we introduce \textbf{Re}al \textbf{D}ocument \textbf{E}mbeddings from \textbf{R}elevance \textbf{F}eedback (ReDE-RF). Inspired by relevance feedback, ReDE-RF proposes to re-frame hypothetical document generation as a relevance estimation task, using an LLM to select which documents should be used for nearest neighbor search. Through this re-framing, the LLM no longer needs domain-specific knowledge but only needs to judge what is relevant. Additionally, relevance estimation only requires the LLM to output a single token, thereby improving search latency. Our experiments show that ReDE-RF consistently surpasses state-of-the-art zero-shot dense retrieval methods across a wide range of low-resource retrieval datasets while also making significant improvements in latency per-query.

%% file: sections/1_intro.tex
\section{Introduction}

\begin{figure*}[t!]
  \includegraphics[width=\textwidth]{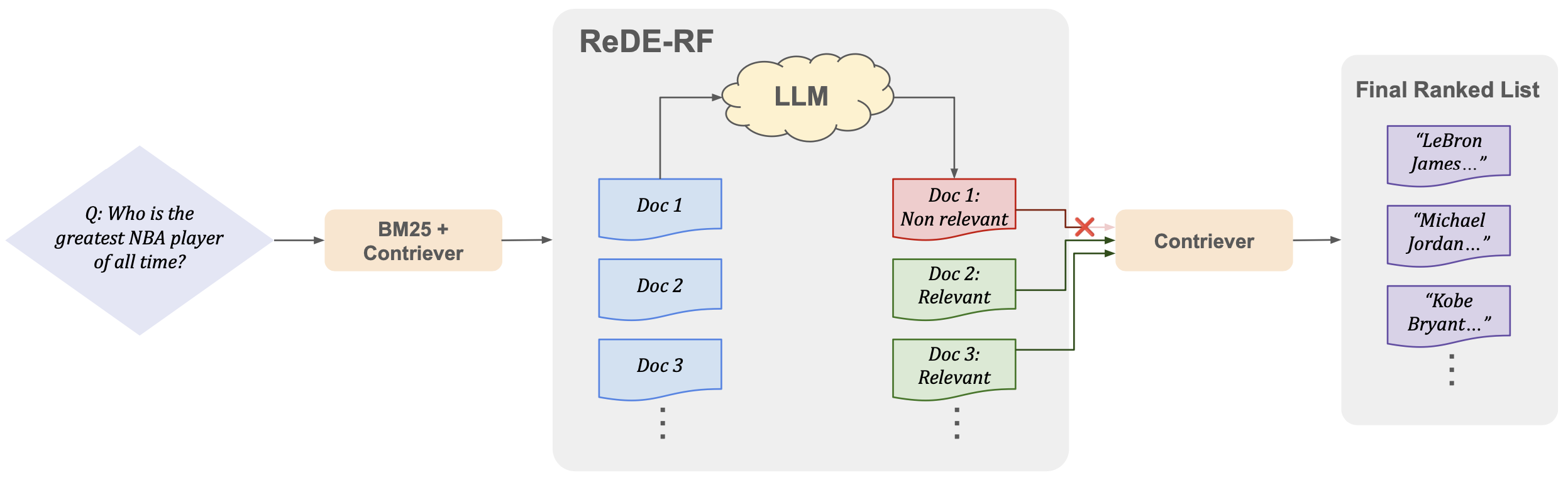}
  \caption{An illustration of the ReDE-RF approach. }
  \label{fig:rede}
\end{figure*}

Information Retrieval (IR) aims to identify relevant documents from a large collection of text given a user's information needs. With recent advancements in transformer-based language models, dense retrieval techniques \cite{karpukhin-etal-2020-dense, xiong2020approximate, qu2020rocketqa} -- which map queries and documents to a shared semantic embedding space that captures relevance patterns -- have demonstrated significant success compared to traditional retrieval approaches based on exact-term matching, such as BM25 \cite{robertson2009probabilistic}. Despite great performance, it remains difficult to build dense retrieval systems in settings that do not have large amounts of dedicated training data \cite{thakur2021beir}.

Recent work has explored improving unsupervised dense retrieval systems, such as Contriever \cite{izacard2021unsupervised}, by leveraging Large Language Models (LLMs) to enrich the query embedding for nearest neighbor search. HyDE \cite{gao-etal-2023-precise}, for example, prompts an LLM to generate hypothetical documents that are used to search for the closest \textit{real} documents. This casts dense retrieval as a document similarity task, which aligns well with pre-training techniques of unsupervised dense retrieval methods \cite{gao-etal-2023-precise, izacard2021unsupervised}.  While HyDE demonstrates strong zero-shot\footnote{In this paper, zero-shot and unsupervised are used without distinction.} performance, it is highly reliant on the LLMs' parametric knowledge, which can be a barrier in deployment for out-of-domain corpora settings (e.g., proprietary documents). A potential solution could be to leverage top-retrieved documents as context when prompting the LLM to generate hypothetical documents \cite{shen2024retrieval}.  However, this increases search latency due to the longer input context. Additionally, even with better prompt context, hypothetical documents generated by LLMs remain susceptible to common issues such as overlooking or ignoring the provided information  \cite{zhou-etal-2023-context, shi2023trusting, liu2024lost, simhi2024constructing}. 

To address these challenges, we propose re-framing the task as \textit{relevance estimation} rather than \textit{hypothetical document generation}. Drawing inspiration from relevance feedback, we introduce \textbf{Re}al \textbf{D}ocument \textbf{E}mbeddings from \textbf{R}elevance \textbf{F}eedback (ReDE-RF).  ReDE-RF first retrieves an initial set of documents from a fully unsupervised hybrid sparse-dense retrieval system and prompts an LLM to mark the returned documents as relevant or non-relevant. Then, given the set of relevant documents, ReDE-RF fetches the document embeddings -- which are precomputed offline -- from the dense index and generates an updated query vector. When updating the query representation with LLM relevance feedback, the new representation is based strictly on \textit{real} documents from the corpus; the LLM does not generate any content that is used to refine the query representation.  Importantly, we also note that ReDE-RF's goal is similar to that of \citet{gao-etal-2023-precise}: we want to develop a full zero-shot dense retrieval pipeline that requires no relevance supervision. 

The core motivation behind our approach is simple: if we can easily access top retrieved documents, we do not need to exclusively rely on LLMs to generate hypothetical documents. First, employing an LLM to generate a hypothetical document for every query is inefficient and introduces unnecessary latency costs. Second, we argue that the task of generating a hypothetical document is highly complex and requires the LLM to (1) already memorize the domain-specific knowledge relevant to the query and (2) replicate the structure of a relevant document. In contrast, knowing what is relevant is a much simpler task. Furthermore, when making use of real documents, we guarantee that the content used to refine the query representation is inherently grounded in the corpus, enabling our method to more seamlessly generalize across different domains. 

We empirically evaluate ReDE-RF on a wide range of retrieval tasks. Our findings reveal that for low-resource tasks, ReDE-RF surpasses zero-shot dense retrieval methods that use LLMs for hypothetical document generation by up to 6\% when LLMs are prompted to generate a hypothetical document with top-retrieved documents as context and 14\% when prompted without. Furthermore, ReDE-RF reduces retrieval latency by as much as 7.5-11.2$\times$ compared to hypothetical document generation with top-retrieved documents as context and 4.4$\times$ without.

Our contributions are summarized as follows:
\begin{itemize}
  \setlength\itemsep{0em}
  \item We propose ReDE-RF, a method that enhances query embeddings for unsupervised dense retrieval systems while addressing key challenges associated with approaches that rely entirely on hypothetical document generation.  
  \item We comprehensively evaluate ReDE-RF on a variety of search tasks and show that ReDE-RF improves upon state-of-the-art zero-shot dense retrieval approaches in low-resource domains while also improving latency.
  \item We demonstrate an approach for \textit{distilling} ReDE-RF to a smaller, more efficient unsupervised dense retriever, DistillReDE. DistillReDE is able to make a 33\% improvement on Contriever while not requiring any update to the Contriever document index or relying on LLMs at inference time. 
\end{itemize}

%% file: sections/2_method.tex
\section{Methodology}

In this section, we first provide a brief overview of HyDE, a method that performs zero-shot dense retrieval through the generation of \textit{hypothetical} documents.
We then describe how we leverage LLMs to perform relevance feedback, a critical component that allows us overcome the challenges of hypothetical documents. Lastly, we describe how we use relevance feedback outputs to update our query representation.

\subsection{Preliminaries: HyDE}

The main challenge in zero-shot dense retrieval is learning query and document embedding functions that capture relevance when human-annotated relevance scores are not available. HyDE \cite{gao-etal-2023-precise} seeks to overcome this by re-casting the task as a document-document similarity task. 
 
Given a query, $q$, HyDE first samples $N$ hypothetical documents $\{\hat{d}_1, \dots, \hat{d}_N\}$ from a generative LLM -- denoted by $\textnormal{LLM}_{\textnormal{DocGen}}$ -- via zero-shot prompting:
\begin{equation}
    \hat{d}_{i}=\textnormal{LLM}_{\textnormal{DocGen}}(q), \quad 1 \leq i \leq N
\end{equation}
where $\textnormal{LLM}_{\textnormal{DocGen}}(q)$ denotes the stochastic output of $\textnormal{LLM}_{\textnormal{DocGen}}$ given $q$. One could optionally provide the top-\textit{k} documents $D = \{d_{1}, d_{2}, \dots, d_{k}\}$ from an unsupervised retrieval system (e.g., BM25) as context for $\textnormal{LLM}_{\textnormal{DocGen}}$: 
\begin{equation}
    \hat{d}_{i}=\textnormal{LLM}_{\textnormal{DocGen}}(D, q), \quad 1 \leq i \leq N
\end{equation}

\noindent
We refer to this as \text{HyDE}$_\text{PRF}$\footnote{We note that this shares similarity with LameR \cite{shen2024retrieval}, but LameR is built on BM25 and has slight implementation differences.}. 

Subsequently, the hypothetical documents,  $\{\hat{d}_1, \dots, \hat{d}_N\}$, are encoded by an unsupervised contrastive encoder, $f$, and averaged to generate an updated query embedding, $\hat{v}_{q_{\textnormal{HyDE}}}$. When generating $\hat{v}_{q_{\textnormal{HyDE}}}$, the original query is also considered. More formally, 
\begin{equation}
    \hat{v}_{q_{\textnormal{HyDE}}} = \frac{1}{N + 1}\ \left ( f(q) + \sum_{i=1}^{N} {f(\hat{d}_i)} \right )
    \label{eq:HyDE}
\end{equation}
$\hat{v}_{q_{\textnormal{HyDE}}}$ is then searched against the corpus embeddings to retrieve the most similar \textit{real} documents. 

Through this two-step process, zero-shot dense retrieval moves from directly modeling query-document similarity to modeling document-document similarity. Without the need for relevance supervision, HyDE is able to out-perform state-of-the-art unsupervised dense retrievers. 

\subsection{ReDE-RF}

Similar to HyDE, ReDE-RF models zero-shot dense retrieval as a document similarity task. Unlike HyDE, the LLM is leveraged for relevance \textit{feedback} rather than 
document \textit{generation}.  ReDE-RF has two key components: (1) relevance feedback with LLMs and (2) updating the query representation. These two components are described in this subsection and illustrated in Figure \ref{fig:rede}. 

\subsubsection{Relevance Feedback with LLMs}  \label{sec:rel_feed} Given a query, we first retrieve the top-\textit{k} documents, $D$, from an unsupervised retrieval system. Subsequently,  we employ zero-shot prompting to score the relevance of given document, $d_{i}$, to the query. Based on the prompt, a generative LLM, denoted by $\textnormal{LLM}_{\textnormal{Rel-Judge}}$, returns a list of the $k^*$ documents classified as relevant $D_{r} = \{d_{r_{1}}, d_{r_{2}}, \dots, d_{r_{k^{*}}}\}$, where $r_i \in \{1,2,\hdots,k^{*}\}$ for $1 \leq i \leq k^*$.  
\begin{figure}[t]
\centering
\includegraphics[width=1.0\linewidth]{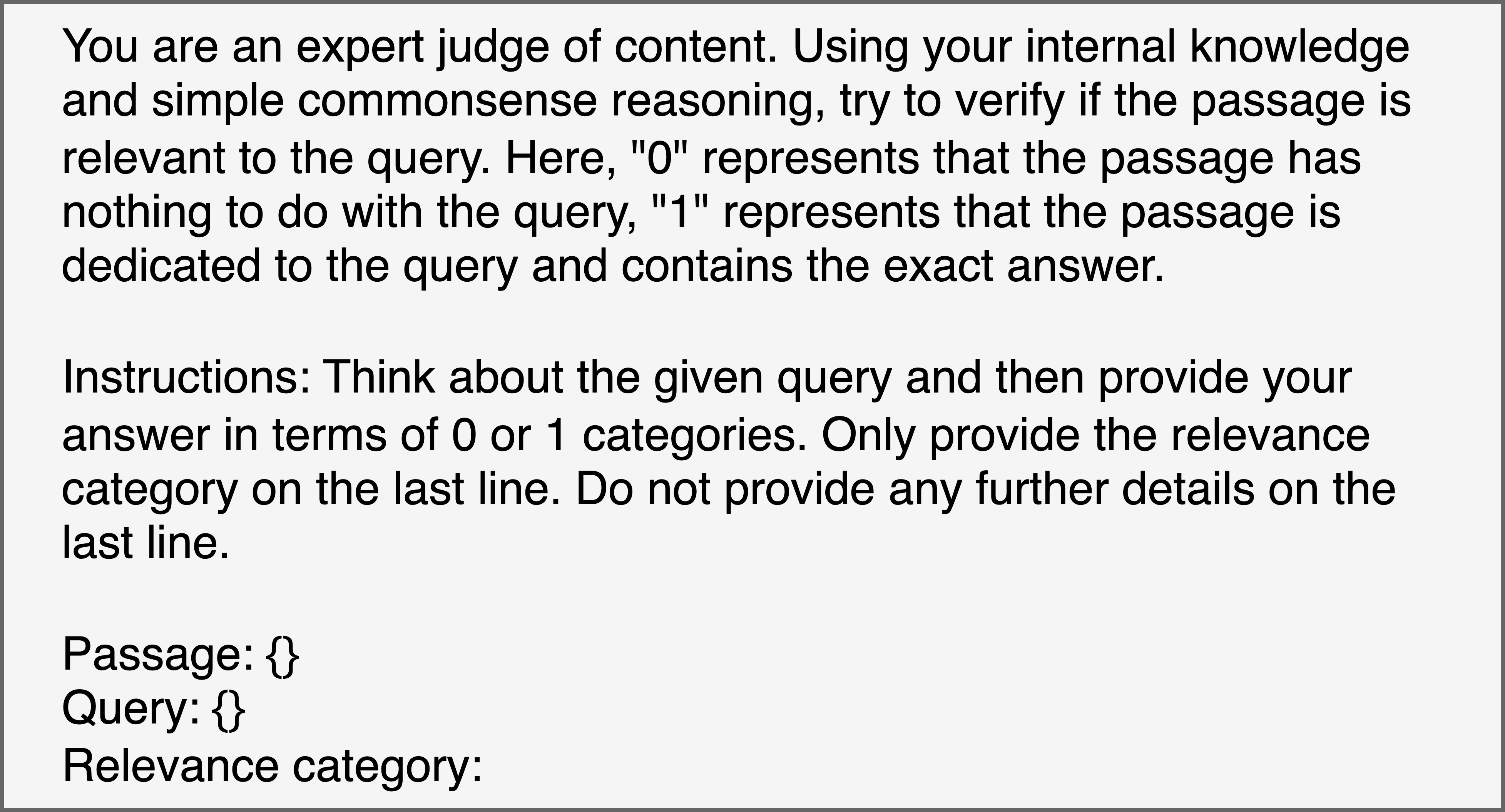}
\caption{ Prompt for relevance feedback, which is a modified version of the prompt used in  \citet{upadhyay2024llms}. \{\} denotes the placeholder for the corresponding text.}
\label{fig:prompt}
\end{figure}
With the recent success of using LLMs for patching up missing relevance judgements, we use a modified version of the prompt from \citet{upadhyay2024llms}, which is shown in Figure \ref{fig:prompt}. 

\subsubsection{Updating the Query Representation} \label{sec:query_rep}
Given the list of documents, $D_{r}$, that $\textnormal{LLM}_{\textnormal{Rel-Judge}}$ deems relevant, we follow Equation \ref{eq:HyDE} to update the query embedding. One key difference between ReDE-RF and HyDE is that $f(d_{r_{i}})$ -- where $1 \leq {i} \leq k^{*}$ -- already exists as it is the embedding of a real document that was pre-computed offline. As such, we denote $C_{E}[d_{r_{i}}]$ as the action of retrieving the embedding for a specified document, $d_{r_{i}}$, from the set of corpus embeddings $C_{E}$. Thus, to update our query:

\begin{equation}
 \hat{v}_{q_{\textnormal{ReDE}}} =  \frac{1}{k^* + 1}\ \left ( f(q) + \sum_{i=1}^{k^*} {C_{E}[d_{r_{i}}]} \right )
\end{equation}

If no relevant documents are found in the top-$k$, i.e., $D_{r} = \emptyset$, a simple option could be to default to the unsupervised contrastive encoder, and just return $f(q)$. However, in these cases we also argue defaulting to hypothetical document generation can be a viable option as it would only hurt latency for difficult queries that the initial retrieval struggles with.  We compare the trade-off between effectiveness and efficiency of these two options in Section \ref{sec:main_results_benchmarks} and \ref{sec:main_results_latency}. 

%% file: sections/3_experiments.tex
\section{Experiments}
\begin{table*}[t!]
  \centering
  \resizebox{\textwidth}{!}{
  \begin{tabular}{l|c|cc|ccccccc|c}
    \toprule
    \multicolumn{1}{c}{} & \multicolumn{1}{c|}{} & \multicolumn{2}{c|}{\bf High Resource} &  \multicolumn{7}{c}{\bf Low Resource (BEIR)}\\
    \cmidrule(lr){3-4}\cmidrule(lr){5-12}
    \multicolumn{1}{l|}{\textbf{Model}} & \multicolumn{1}{c|}{\textbf{Init. Retrieval}} & DL19 & DL20 & News & Covid & FiQA &  SciFact & DBPedia & NFCorpus &  Robust04 & BEIR (Avg.) \\
    \midrule
    BM25 & N/A & 50.6 & 48.0 & 39.5 & 59.5 & 23.6 & 67.9 & 31.8 & 32.2 & 40.7 & 42.2 \\
    Contriever & N/A  & 44.5 &	42.1 &	34.8 &	27.3 &	24.5 &	64.9 &	29.2 &	31.7 &	31.6 & 34.9  \\
    Hybrid (BM25 + Contriever) & N/A  & 52.9 &	50.9 &	42.2 &	52.9 &	28.4 &	71.6 &	34.8 &	33.7 &	43.0 &	43.8  \\
     $\textnormal{Contriever}_{\textnormal{AvgPRF}}$ & & & & & & & & & & & \\
    \quad  PRF-Depth: 3 & Hybrid (3) &  52.1 &	46.4 &	43.4 &	48.1 &	22.8 &	59.9 &	31.9 &	32.3 &	38.6 &	39.6 \\
    \quad  PRF-Depth: 20 & Hybrid (20) &  49.2 &	46.2 &	39.9 &	51.3 &	12.0 &	36.9 &	28.6 &	21.7 &	39.2 & 32.8\\
    \midrule
    \multicolumn{1}{l}{\textbf{Zero-Shot Dense Retrieval}} \\
    \midrule
    PromptReps (Llama3-8B-I) & N/A & - & - & - & 59.5 & 27.1 & 52.7 & 31.5 & 29.6 & - & - \\
    HyDE (Mistral-7B-Instruct) & N/A & 57.8 &	53.9 &	44.0 &	56.9 &	21.6 &	65.1 &	35.3 &	27.7 &	41.5 &	41.7 \\
    \text{HyDE}$_\text{PRF}$ (Mistral-7B-Instruct) & Hybrid (20) & \textbf{63.5} &	\textbf{62.0} &	46.9 &	59.1 &	28.3 &	64.5 &	35.2 &	35.0 &	45.6 &	44.9\\[-1em]
    \multicolumn{12}{c}{\tikz[baseline]{\draw[dashed] (0,-0.1cm) -- (24cm,-0.1cm);}}\\
    ReDE-RF (Ours)  & & & & & & & & & & & \\
    \quad Default: Contriever & Hybrid (20) & 60.3 & 59.4 & 46.1 &	\textbf{65.6} &	28.2 &	\textbf{67.4} &	37.0 &	34.8 &	49.8 &	47.0  \\
    \quad Default: \text{HyDE}$_\text{PRF}$ & Hybrid (20) & 62.8 &	60.4 &	\textbf{47.1} &	65.6 &	\textbf{29.3} &	 66.9 &	\textbf{37.6} &	\textbf{35.5} &	\textbf{51.7} &	\textbf{47.7} \\
    \midrule
    \multicolumn{1}{l}{\textbf{Supervised Dense Retrieval}} \\
    \midrule
    DPR & N/A & 62.2 & \textbf{65.3} & 16.1 & 33.2 & 11.2 & 31.8 & 26.3 & 18.9 & 25.2 & 23.2 \\
    ANCE & N/A & \textbf{64.5} & 64.6 & 38.2 & \textbf{65.4} & 29.5 & 50.7 & 28.1 & 23.7 & 39.2 & 39.3  \\
    $\textnormal{Contriever}^{\textnormal{FT}}$ & N/A & 62.1 & 63.2 & \textbf{42.8} & 59.6 & \textbf{32.9} & \textbf{67.7} & \textbf{41.3} & \textbf{32.8} & \textbf{47.3} & \textbf{46.3} \\
    \bottomrule
  \end{tabular}}
  \caption{Results (\textbf{NDCG@10)} on TREC and BEIR. We report the mean NDCG@10 across \textit{three} runs for HyDE, \text{HyDE}$_\text{PRF}$, and ReDE-RF (Default: \text{HyDE}$_\text{PRF}$). The average standard deviation across all datasets for HyDE, \text{HyDE}$_\text{PRF}$ and ReDE-RF (Default:\text{HyDE}$_\text{PRF}$)  was $\approx$ 0.4\%, $\approx$ 0.5\% and $\approx$ 0.1\% respectively. Exact numbers can be found in Appendix \ref{sec:standard_deviations}.} 
  \label{tab:BEIR}
\end{table*}

\subsection{Setup}
\label{sec:setup}

\textbf{Implementation} \quad ReDE-RF requires an instruction-tuned LLM and a dense retriever. For the instruction-tuned LLM (i.e., $\textnormal{LLM}_{\textnormal{Rel-Judge}}$) we leverage Mistral-7B-Instruct-v0.2 \cite{jiang2023mistral} and for dense retrieval we use the unsupervised Contriever \cite{izacard2021unsupervised}. When prompting $\textnormal{LLM}_{\textnormal{Rel-Judge}}$ for relevance feedback, we truncate the input document to at most 128 tokens and generate a relevance score by applying a softmax on the logits of the “1” and “0” tokens as shown in \citet{nogueira-etal-2020-document}. Only documents that $\textnormal{LLM}_{\textnormal{Rel-Judge}}$ scores as '1' are used for updating the query representation. In cases in which $D_{r} = \emptyset$, we consider two defaults: Contriever and \text{HyDE}$_\text{PRF}$.

To generate an initial document set for $\textnormal{LLM}_{\textnormal{Rel-Judge}}$, we retrieve the top-20 documents from a hybrid, sparse-dense, retrieval model (BM25 + Contriever)\footnote{We provide implementation details in Appendix \ref{sec: hybrid_retrieval_desc}}. Retrieval experiments were performed with Pyserini \cite{lin2021pyserini} and LLM implementations in HuggingFace \cite{wolf2019huggingface}.

\smallskip
\noindent
\textbf{Datasets} \quad  In our experiments, we evaluate on two web search datasets: TREC DL19 \cite{craswell2020overview} and TREC DL20 \cite{craswell2021overview}. We also evaluate on seven low-resource retrieval datasets from BEIR \cite{thakur2021beir}. The tasks include news retrieval (TREC-News, Robust04), financial question answering (FiQA), entity retrieval (DBpedia), biomedical IR (TREC-Covid, NFCorpus), and fact checking (SciFact). For metrics, we report NDCG@10, the offical metric for the TREC and BEIR datasets. 

\smallskip
\noindent
\textbf{Baselines} We first compare ReDE-RF to unsupervised retrievers that do not leverage LLMs:  BM25, Contriever, and a hybrid retrieval model (BM25 + Contriever). We also include a pseudo-relevance feedback (PRF) baseline which averages all top-$k$ initially retrieved documents to update the query representation:  $\textnormal{Contriever}_{\textnormal{AvgPRF}}$ \cite{li2022pseudo}. $\textnormal{Contriever}_{\textnormal{AvgPRF}}$ is equivalent to ReDE-RF if all top-$k$ retrieved documents are considered relevant. For $\textnormal{Contriever}_{\textnormal{AvgPRF}}$, the initially retrieved documents are from the hybrid retrieval model (BM25 + Contriever). 

We then compare ReDE-RF to methods that use LLMs for zero-shot dense retrieval and require no training.  Our main point of comparison for ReDE-RF is HyDE and \text{HyDE}$_\text{PRF}$. For \text{HyDE}$_\text{PRF}$ we prompt the LLM using the top-20 initially retrieved documents from the same hybrid retrieval system as ReDE-RF. We also compare ReDE-RF to the dense retrieval version of PromptReps \cite{zhuang2024promptreps}, which generates query and document representations by prompting an LLM to generate a single token that describes the text. 

Lastly, we compare against dense retrieval systems that have been fine-tuned with supervised data: DPR \cite{karpukhin-etal-2020-dense},  ANCE \cite{xiong2021approximate}, and $\textnormal{Contriever}^{\textnormal{FT}}$, a fine-tuned version of Contriever. 

\subsection{Results on Benchmarks}
\label{sec:main_results_benchmarks}
Table \ref{tab:BEIR} presents the evaluation results on the TREC and BEIR datasets and reveals several insights: 

(1) ReDE-RF outperforms  $\textnormal{Contriever}_{\textnormal{AvgPRF}}$, which uses all  initially retrieved documents to enhance the query embeddings. This exemplifies that simply leveraging the top-$k$ retrieved documents is not sufficient, and demonstrates the value of leveraging an LLM to filter out non-relevant documents. 

(2) Comparing ReDE-RF to HyDE, we find that using real documents for zero-shot dense retrieval consistently outperforms  hypothetical documents based solely on LLM knowledge (i.e., without top documents as context). 

(3) When incorporating corpus text as a guide for HyDE (i.e., \text{HyDE}$_\text{PRF}$), the performance gap between ReDE-RF and HyDE decreases. However, ReDE-RF still provides substantial improvements in low-resource domains (6.0\% when defaulting to \text{HyDE}$_\text{PRF}$ and 4.6\% when defaulting to Contriever). For high-resource domains — DL19 and DL20 — \text{HyDE}$_\text{PRF}$ yields better results, which we hypothesize is due to the advantages of combining the LLM's parametric knowledge with corpus knowledge in domains the LLM is well-versed. 

(4) As the performance of ReDE-RF (Default: \text{HyDE}$_\text{PRF}$) is equivalent to \text{HyDE}$_\text{PRF}$ for queries that ReDE-RF defaults, the performance gains of ReDE-RF in low-resource domains can be attributed to the benefits of doing nearest-neighbor search in the \textit{real} document embedding space -- that  $\textnormal{LLM}_{\textnormal{Rel-Judge}}$ deemed relevant -- versus the \textit{hypothetical} document space generated by  $\textnormal{LLM}_{\textnormal{DocGen}}$ 

(5) While ReDE-RF remains competitive with fine-tuned dense retrieval systems (DPR, ANCE, and $\textnormal{Contriever}^{\textnormal{FT}}$) in high-resource domains, in low-resource domains, ReDE-RF outperforms DPR and ANCE on all but one dataset, and surpasses $\textnormal{Contriever}^{\textnormal{FT}}$ on four of the seven low-resource datasets.

\subsection{Comparing Latencies}

\label{sec:main_results_latency}

\begin{figure}[t]
\centering
\includegraphics[width=1.0\linewidth]{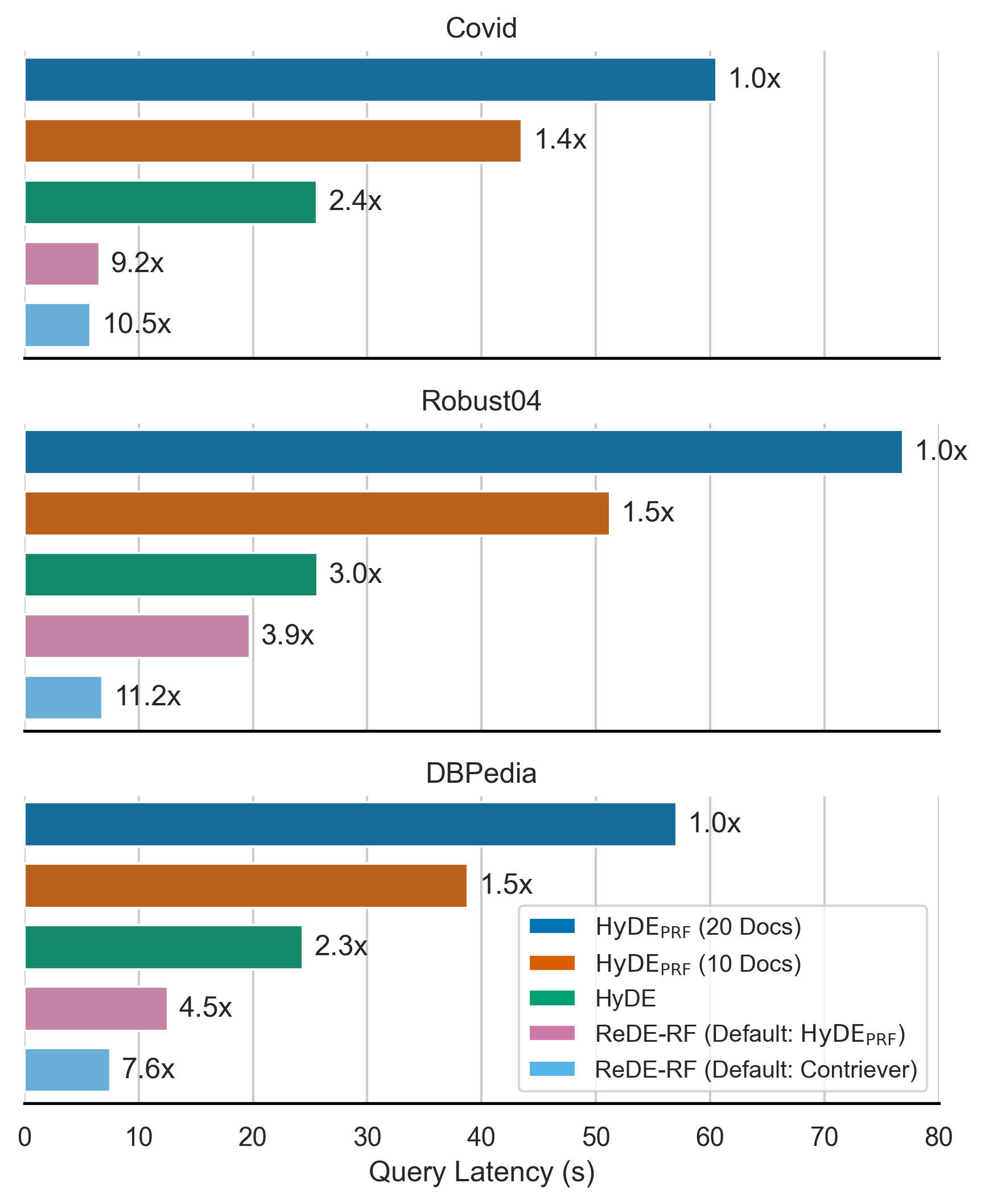}
\caption{Latency per query for \text{HyDE}$_\text{PRF}$, HyDE and ReDE-RF. Speedup is relative to slowest method (\text{HyDE}$_\text{PRF}$).}
\label{fig:latency_main}
\end{figure}

In Figure \ref{fig:latency_main},  we empirically compare the average query latency for \text{HyDE}$_\text{PRF}$, HyDE and ReDE-RF. We also include \text{HyDE}$_\text{PRF}$ with 10 initially retrieved documents -- \text{HyDE}$_\text{PRF}$ (10 Docs) -- as an additional comparison\footnote{NDCG@10 results can be found in Appendix \ref{sec:hyde_prf_less_docs}}.  All experiments were run on one A100 GPU and measure the time from input query to retrieval of results.  

Comparing the latencies across systems, we find that ReDE-RF (Default: Contriever)  consistently reduces latency compared to HyDE and \text{HyDE}$_\text{PRF}$. Specifically, on average, ReDE-RF (Default: Contriever) is 3.8$\times$ faster than HyDE and 6.7 to 9.7$\times$ faster than \text{HyDE}$_\text{PRF}$, depending on whether 10 or 20 documents are used as context. This finding is true even when ReDE-RF defaults to \text{HyDE}$_\text{PRF}$, improving latency by 2.4$\times$ compared to HyDE and and 4.1 to 5.9$\times$ compared to \text{HyDE}$_\text{PRF}$. These results  confirm our hypothesis that leveraging hypothetical document generation for every query introduces unnecessary latency costs and that it is possible to improve performance while also improving efficiency.   

For ReDE-RF, Table \ref{tab:BEIR} showed that defaulting to \text{HyDE}$_\text{PRF}$ can provide a performance boost as compared to defaulting to Contriever. However, these improvements come with a higher latency. Ideally, ReDE-RF can achieve the performance of  ReDE-RF (Default: \text{HyDE}$_\text{PRF}$) while fully removing the need for generating hypothetical documents. We investigate this in Section \ref{sec:distill_rede}. 

%% file: sections/4_ablation.tex
\section{Ablation Study on ReDE-RF}

There are many design decisions that one can make when implementing ReDE-RF. In this section, we study the effects of these different choices on ReDE-RF's performance. As some approaches may default more frequently than others, the result will be affected by how strong the chosen default is. Thus, in the ablation study, we choose no default: return no results for the query if $k^* = 0$ (yielding an NDCG@10 of 0) to limit our study solely to the relevance feedback portion of ReDE-RF. We refer to this as ReDE-RF (No Default.) 

\label{sec:Initial_Retrieval_Abalation}

\smallskip
\noindent
\textbf{Effect of Initial Retrieval Method} \quad How does the initial retriever effect ReDE-RF accuracy? The results for this experiment are in Table \ref{tab:initial_ret}. Feeding documents using hybrid retrieval consistently improves results compared to only sparse or only dense retrieval. As ReDE-RF is highly dependent on the initial retrieval, these results suggest that leveraging multiple unsupervised retrievers can improve performance.

\begin{table}[t!]
  \centering
  \resizebox{1.0\linewidth}{!}{
  \begin{tabular}{l|cccc}
    \toprule
    \textbf{Method} & \bf DL20 & \bf Covid  & \bf FiQA \\
    \midrule
    \textbf{Init. Retrieval: BM25 } \\
    ReDE-RF (No Default) &  56.7 & 65.0  & 20.6 \\
    \midrule
    \textbf{Init. Retrieval: Contriever} \\
    ReDE-RF (No Default) & 52.4  & 37.3 & 21.6 \\
    \midrule
    \textbf{Init. Retrieval: Hybrid} \\
    ReDE-RF (No Default) & 59.0 & 65.5 & 22.9 \\
    \bottomrule
  \end{tabular} }
  \caption{Impact of initial retriever on NDCG@10.}
  \label{tab:initial_ret}
\end{table}

\smallskip
\noindent
\textbf{Impact of $k^*$} \quad In Table \ref{tab:retrieved_res} we study how the number of relevant documents ($k^*$ as defined in \ref{sec:rel_feed}) used to update the query representation influences the accuracy of ReDE-RF. Accuracy generally increases up to and stabilizes around $k^* = 10$. We hypothesize this may be due to the increased number of potential false positives returned as the number of relevant documents increases, which may push ReDE-RF's query embedding further away from the true positives. 

\begin{table}[t!]
  \centering
  \small
  \begin{tabular}{l|cccc}
    \toprule
    & \bf DL20 & \bf News & \bf DBpedia \\
    \midrule
    \textit{Max $k^*$} \\
    1 &  54.2 & 35.8 & 31.5 \\
    5 &  58.5 & 40.8 & 35.0 \\
    10 & 60.0 & 41.1 & 36.1 \\
    20 & 59.0 & 41.2 & 35.4 \\ 
    \bottomrule
  \end{tabular} 
  \caption{Impact of the number of documents used to update the ReDE-RF (No Default) query embedding.}
  \label{tab:retrieved_res}
\end{table}

\smallskip
\noindent
\textbf{Effect of Relevance Feedback Model} \quad In Table \ref{tab:model_impact} we investigate ReDE-RF with different LLMs. We generally find similar trends across model families: as the model size increases, performance improves. Generally, smaller models are competitive with their larger counterparts, demonstrating the potential for using a smaller LLM for relevance feedback at the benefit of faster inference times.     

\begin{table}[t!]
  \centering
  \small
  \resizebox{1.0\linewidth}{!}{
  \begin{tabular}{l|ccc|c}
    \toprule
    \textbf{Method} & \bf DL19  & \bf DL20 \\
    \midrule
    ReDE-RF (No Default)&  \\
    \quad  w/ Mistral-7B-Instruct & 59.1 & 59.0   \\
    \quad  w/ Mixtral-8x7B-Instruct & 62.1 & 58.7 \\[-1.0em]
    \multicolumn{5}{c}{\tikz[baseline]{\draw[dashed] (0,-0.06cm) -- (6cm,-0.06cm);}} \\
    \quad w/ Gemma-2-2B-it & 58.9 &	56.7 \\
    \quad w/ Gemma-2-9B-it & 62.0 &	58.9 \\[-1.0em]
    \multicolumn{5}{c}{\tikz[baseline]{\draw[dashed] (0,-0.06cm) -- (6cm,-0.06cm);}} \\
    \quad w/ Llama-3.2-3B-I & 56.0 & 54.1\\
    \quad w/ Llama-3.1-8B-I & 57.2 & 56.1 \\
    \bottomrule
    \end{tabular} 
    }
  \caption{Impact of Relevance Feedback model on NDCG@10. For Mixtral-8x7B-Instruct, we perform inference with 4-bit quantization.}
  \label{tab:model_impact}
\end{table}

\noindent
\textbf{Prompt Variations} \quad In Table \ref{tab:prompt_variations} we study how different prompts impact ReDE-RF's retrieval effectiveness. The results show that, on average, performance across prompts is similar. The only prompt that performs worse is RG-YN, which asks \textit{"For the following query and document, judge whether they are relevant. Output “Yes” or “No”}. We hypothesize that the drop in performance stems from the prompt not giving a clear definition of what relevance is (which the other prompts do more clearly, e.g., \textit{"the passage answers the query"}). To test this, we augment the prompt with the description relevance from Figure \ref{fig:prompt}  (RG-YN$*$). The results show that this simple augmentation improves performance of RG-YN by 4.6\%, on average. This finding hints that when creating prompts for ReDE-RF it is important the prompt includes a clear definition of what should be classified as relevant versus not. 

\begin{table}[t!]
  \centering
  \resizebox{1.0\linewidth}{!}{
  \begin{tabular}{l|ccc|c}
    \toprule
    \textbf{Prompt} & \bf DL19  & \bf DL20 \\
    \midrule
    Figure \ref{fig:prompt} & 59.1 & 59.0  \\
    pointwise.yes\_no \cite{zhuang2024setwise} & 61.4 & 56.1 \\
    RG-YN \cite{zhuang2024beyond} & 57.8 & 51.8 \\
    RG-YN$*$  & 59.8 & 54.9 \\
    \citet{thomas2024large} & 61.4 & 56.9 \\
    \bottomrule
    \end{tabular} }
  \caption{Impact of different prompts on ReDE-RF (No Default). For \citet{thomas2024large}, we make the relevance options binary.  Prompts are in Appendix \ref{sec:rede_prompt_variations}.}
  \label{tab:prompt_variations}
\end{table}

%% file: sections/5_distill_rede.tex
\section{Can we Distill ReDE-RF?} 

\label{sec:distill_rede}

As noted in Section \ref{sec:main_results_latency}, ReDE-RF improves latency per query as compared to HyDE and \text{HyDE}$_\text{PRF}$. However, if ReDE-RF is implemented with \text{HyDE}$_\text{PRF}$ as its default, it still occasionally needs to default to hypothetical document generation if no relevant documents are found, thus making it costly for certain queries. In this section, we explore if we can improve the latency of ReDE-RF without trading off accuracy. With this in mind, we aim to answer two questions: 1) Can we \textit{distill} ReDE-RF's performance to Contriever (DistillReDE)?  2) Can using DistillReDE in tandem with ReDE-RF remove the need for defaulting to \text{HyDE}$_\text{PRF}$ while matching the performance of ReDE-RF (Default: \text{HyDE}$_\text{PRF}$)?

\begin{figure}[t!]
\centering
\includegraphics[width=1.0\linewidth]{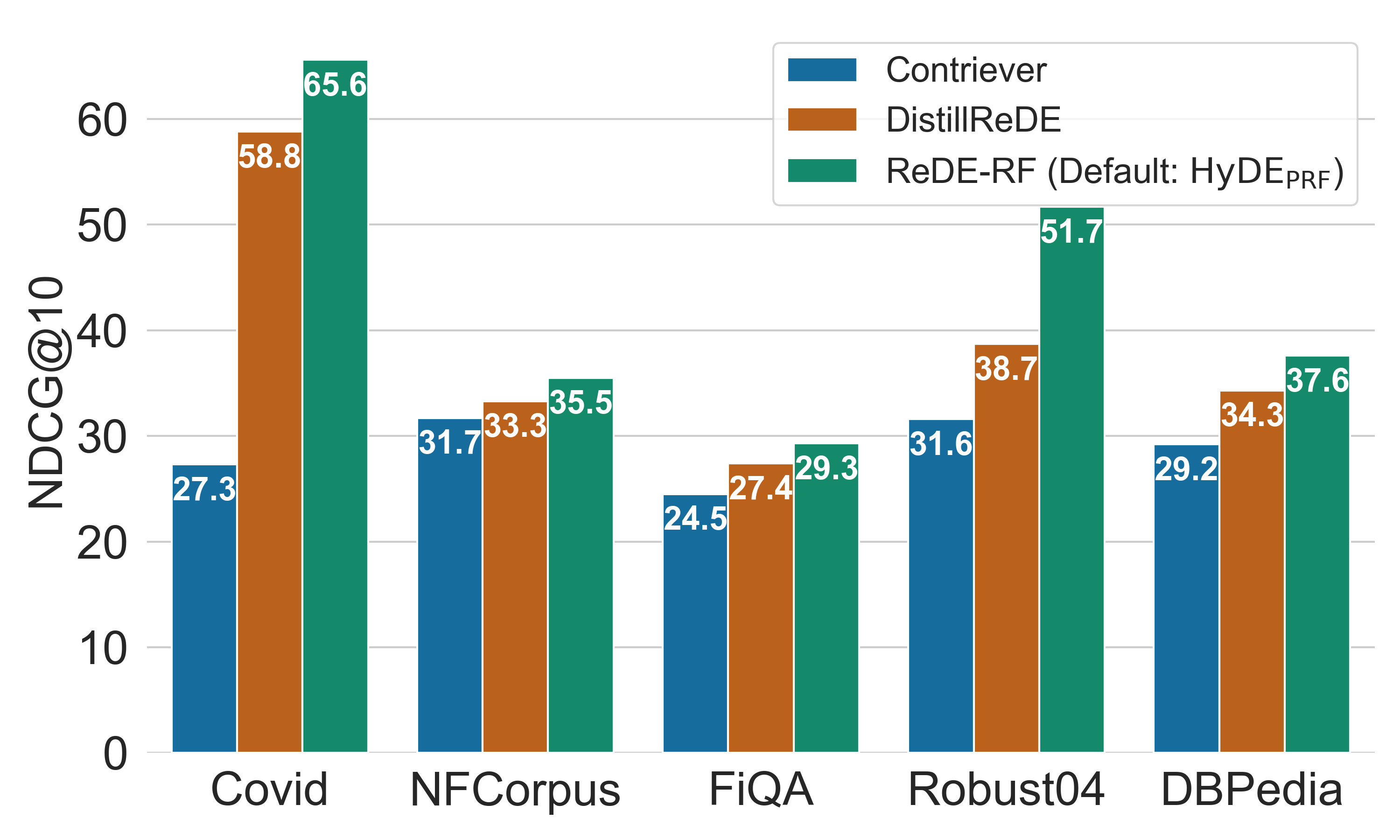}
\caption{Comparison of DistillReDE to Contriever and ReDE-RF (Default: \text{HyDE}$_\text{PRF}$).}
\label{fig:distill_rede}
\end{figure}

\smallskip
\noindent
\textbf{Distilling ReDE-RF} \quad We aim to explore whether ReDE-RF can be distilled to a student Contriever model, DistillReDE. Since ReDE-RF’s embeddings are an average of the Contriever document embeddings, one advantage is that the student model can be trained without the need to re-index the corpus. To generate the training set, we first run ReDE-RF offline using LLM-generated synthetic queries and treat the corresponding ReDE-RF embeddings as the target representation. For training,  we follow the framework from \citet{pimpalkhute2024softqe}:  we optimize a combination of MSE loss and contrastive loss with in-batch random negatives. See Appendix \ref{sec: distill_rede_training} for training details. 

The results, shown in Figure \ref{fig:distill_rede}, indicate that DistillReDE can achieve significant improvements on Contriever, narrowing its performance gap with ReDE-RF (Default: \text{HyDE}$_\text{PRF}$)  while removing the need for LLMs at inference time. 

\smallskip
\noindent
\textbf{ReDE-RF with DistillReDE} \quad We next explore the possible advantages of leveraging DistillReDE as a drop-in replacement for Contriever in the ReDE-RF system. Table \ref{tab:distill_rede_w_rede} shows the results of this experiment. When fully leveraging DistillReDE as the initial retriever and ReDE-RF default (row 3), performance is very competitive compared to ReDE-RF when defaulting to \text{HyDE}$_\text{PRF}$ (row 2). When performing initial retrieval using a hybrid (BM25 + DistillReDE) initial retriever (row 4), performance improves compared to row 3 and increases slightly over row 2, while being significantly less costly (as we remove any need for hypothetical document generation). These results demonstrate that with a simple offline training scheme, ReDE-RF can match the performance of ReDE-RF (Default: \text{HyDE}$_\text{PRF}$) while fully removing the need for hypothetical document generation. 

\begin{table*}[t!]
  \centering
  \small
  \begin{tabular}{l|c|ccccc}
    \toprule
     \bf Default & \bf Init. Retrieval & \bf Covid & \bf NFCorpus & \bf FiQA & \bf Robust04 & \bf DBPedia \\
    \midrule
    Contriever & Hybrid & 65.6 & 34.8 & 28.2 & 49.8 & 37.0 \\
    HyDE$_\text{PRF}$ & Hybrid & 65.6 & 35.5 & 29.3 & \textbf{51.7} & 37.6 \\
    \midrule
    DistillReDE & DistillReDE & 63.8 & 35.6 & 30.2 & 	47.9 & 	37.8 \\
    DistillReDE & Hybrid$^*$ & \textbf{66.3} &	\textbf{35.8} &	\textbf{30.9} &	49.2 &  \textbf{38.4} \\
    \bottomrule
  \end{tabular} 
  \caption{NDCG@10 of ReDE-RF when implemented with DistillReDE. Hybrid$^*$ is a hybrid system that combines results from BM25 and DistillReDE. Hybrid is BM25 + Contriever, as in Table \ref{tab:BEIR}.}
  \label{tab:distill_rede_w_rede}
\end{table*}

%% file: sections/6_rede_v_passage_reranking.tex
\begin{table} 
  \centering
  \resizebox{1.0\linewidth}{!}{
  \begin{tabular}{l|ccc}
    \toprule
     & \bf Covid & \bf DL19 & \bf News \\
     \textbf{Method} & NDCG@10/20 & NDCG@10/20  & NDCG@10/20 \\
    \midrule
    \textbf{Mistral-7B-Instruct} \\
    ReDE-RF & \underline{65.6} / \underline{57.9} & \underline{\textbf{62.8}} / \underline{\textbf{60.3}} & \underline{\textbf{47.1}} / \underline{\textbf{43.8}} \\
    Hybrid + PR & 63.6 / 49.6 & 60.0 / 53.7 & 45.6 / 42.5 \\[-1.0em]
    \multicolumn{4}{c}{\tikz[baseline]{\draw[dashed] (0,-0.06cm) -- (11cm,-0.06cm);}} \\
    ReDE-RF + PR & \textbf{68.8} / \textbf{58.9} & 62.6 / 59.9 &  45.8 / 43.2 \\
    \midrule
    \textbf{Gemma-2-9B-it} \\
    ReDE-RF  & \underline{67.9} / \underline{60.0} &	62.0 /	\underline{61.3} &	\underline{\textbf{46.6}} /	\underline{\textbf{42.8}} \\
    Hybrid + PR  & 65.0 /	50.8 &	\underline{63.7} /	55.8 &	46.3 /	41.9 \\[-1.0em]
    \multicolumn{4}{c}{\tikz[baseline]{\draw[dashed] (0,-0.06cm) -- (11cm,-0.06cm);}} \\
    ReDE-RF + PR   & \textbf{72.3} / \textbf{61.2} &	\textbf{70.7} / \textbf{64.8} &	43.4 /	41.5  \\
    \midrule
    \textbf{Llama-3.1-8B-I} \\
    ReDE-RF & 65.7 /	\underline{58.7} &	59.0 /	\underline{59.0} &	\underline{\textbf{49.9}} /	\underline{\textbf{45.7}} \\
    Hybrid + PR  & \underline{67.1} /	51.8 &	\underline{66.8} /	57.4 &	47.2 /	42.8 \\[-1.0em]
    \multicolumn{4}{c}{\tikz[baseline]{\draw[dashed] (0,-0.06cm) -- (11cm,-0.06cm);}} \\
    ReDE-RF + PR &  \textbf{72.9} /	\textbf{61.1} &	\textbf{70.1} /	\textbf{64.3} &	48.6 /	45.4 \\
    \bottomrule
  \end{tabular}
  }
  \caption{Comparing ReDE-RF (Default: \text{HyDE}$_\text{PRF}$) to pointwise reranking (PR). ReDE-RF + PR reranks the top-20 passages returned from ReDE-RF. \textbf{Bold} denotes best overall system. \underline{Underline} denotes best between ReDE-RF and Hybrid + PR.}
  \label{tab:rede_v_reranking}
\end{table}

\section{ReDE-RF vs. Pointwise Reranking}

The  $\textnormal{LLM}_{\textnormal{Rel-Judge}}$ component of ReDE-RF (discussed in \ref{sec:rel_feed}) is similar to LLM-based pointwise re-rankers \cite{zhuang2024setwise}. In this subsection, we ask: What benefits do we achieve by feeding relevant documents to improve the query representation -- as described in \ref{sec:query_rep} -- versus simply re-ordering the initial retrieval based on the logits from $\textnormal{LLM}_{\textnormal{Rel-Judge}}$? To answer this, we focus on comparing pointwise re-ranking to ReDE-RF (Default: \text{HyDE}$_\text{PRF}$) in equal settings: Both systems have access to the top-20 passages from a hybrid (BM25 + Contriever) retriever and employ the same prompt as shown in Figure \ref{fig:prompt}. Note, for the rest of this section we refer to ReDE-RF (Default: \text{HyDE}$_\text{PRF}$) as ReDE-RF.

In Table \ref{tab:rede_v_reranking} we present the results of this experiment across three backbone LLMs. Based on the results, we can make the following observations: (1) When comparing NDCG@10, ReDE-RF and pointwise re-ranking are generally on par -- outside of Llama-3.1-8B-I on DL19. (2) ReDE-RF consistently outperforms pointwise re-ranking in terms of NDCG@20 by large amounts. This demonstrates that ReDE-RF's improvements extend beyond the top-ranked results and is not confined to the initial retrieval. (3) Besides the evaluation on the TREC News dataset -- where it appears pointwise re-ranking is not well calibrated -- re-ranking and ReDE-RF are generally \textit{complementary}. ReDE-RF + PR outperforms Hybrid + PR eight out of nine times (six out of six if excluding TREC News) and outperforms ReDE-RF five out of nine times (five out of six if excluding TREC News).  

These results exemplify the difference in the roles of passage re-ranking and ReDE-RF. While passage re-ranking primarily enhances the ordering of the top-$k$ passages, ReDE-RF improves the overall quality of candidates from the first-stage retrieval.

%% file: sections/7_related_work.tex
\section{Related Work}
\noindent
\textbf{Query Expansion with LLMs} \quad GAR \cite{mao-etal-2021-generation} was among the first methods to demonstrate the effectiveness of LLMs for query expansion by training an LLM to expand queries through the generation of relevant contexts, such as the target answer or answer sentence. Recent work has looked into leveraging LLMs to generate query expansions via zero or few-shot prompting. This has been explored in contexts where the LLM generates hypothetical documents that can be used to augment the query \cite{gao-etal-2023-precise, wang-etal-2023-query2doc, jagerman2023query, lei-etal-2024-corpus, mackie2023generative, shen2024retrieval}. While some of these works generate hypothetical texts given \textit{real} documents (e.g., \citet{jagerman2023query, lei-etal-2024-corpus,  shen2024retrieval}) they still rely on LLM-generated content to augment the query. 

Other work has looked into improving the outputs of LLM query expansions through query re-ranking \cite{chuang-etal-2023-expand} or further training an LLM with preferences from the target retrieval systems \cite{yoon2024ask}.  

\smallskip

\noindent
\textbf{Zero-Shot Dense Retrieval} \quad  With advancements in deep learning, IR systems moved away from representations based on exact-term matching to dense vector representations generated from transformer language models \cite{lin2022pretrained}, such as BERT \cite{devlin-etal-2019-bert}.  However, learning these representations typically requires large, labeled datasets. As such, researchers have looked into methods for learning dense representations without manually labeled data through techniques such as using synthetic data \cite{izacard2021unsupervised, wang2023improving, sachan2023questions, lee2024gecko, dai2022promptagator}, addressing architecture limitations to improve zero-shot decoder-LLM embeddings \cite{springer2024repetition}, or leveraging LLMs to generate outputs that can be used to improve representations for zero-shot dense retrieval \cite{gao-etal-2023-precise, zhuang2024promptreps}. 

%% file: sections/8_conclusion.tex
\section{Conclusion}

We introduce ReDE-RF, a zero-shot dense retrieval method that addresses key challenges associated with approaches that rely entirely on hypothetical document generation. Through extensive experiments, we show that ReDE-RF improves upon state-of-the-art zero-shot dense retrieval approaches in low-resource domains, while also lowering latency compared to techniques that rely only on hypothetical document generation. Further analysis shows that ReDE-RF can be easily distilled to a smaller, more efficient unsupervised dense retriever, DistillReDE, removing any reliance on LLMs at inference time.  In summary, ReDE-RF presents an approach that achieves the benefits of casting zero-shot dense retrieval as a document similarity task while  being more efficient and domain-agnostic.

%% file: sections/limitations.tex
\section*{Limitations}

A limitation of ReDE-RF is its reliance on retrieved results from first-stage retrievers. If an initial retriever provides a poor set of results, performance gains will not be as apparent as no relevant documents can be used to update the query embedding. This in turn makes ReDE-RF equivalent to Contriever or HyDE, depending on what default the user leverages. How to make ReDE-RF less reliant on retrieved results from unsupervised first-stage retrievers is a question worth exploring in future work. Another simple improvement could be leveraging a rules-based approach that keeps assessing retrieved documents if none of the top results are deemed relevant. We do note that this would likely increase latency. 

Another limitation is that while ReDE-RF seeks to minimize reliance on LLM-generated outputs, it does still depend on an LLM to be accurate in its relevance feedback.  If the LLM provides inaccurate relevance assessments during the relevance feedback stage, it can further harm the query representation. Lastly, while ReDE-RF improves latency compared to previous approaches based on hypothetical document generation, the latency is still slower than approaches that do not rely on LLMs at inference time. However, we demonstrated the potential for doing offline training as a way to mitigate this. Additionally, as 
LLMs advance, we may see improvements in the efficiency of LLM inference. Techniques such as flash-attention \cite{dao2022flashattention}, can also significantly decrease the inference time of LLMs.

\section*{Ethics Statement}

 While the ultimate goal of our work is to minimize reliance on LLM generated output, we do recognize that our system does still rely on LLMs, which means that there is a risk that the LLM can produce biased, harmful, or offensive output. To mitigate this, we limit our LLM to only generate one token, which we hope can eliminate this risk. Additionally, our dense retrieval system is based on pre-trained language models which can potentially produce retrieval results that contain human biases. 

Our research solely uses publicly available datasets, and no personal information is collected. All datasets and models are used in accordance with its intended use and licenses. Our method is designed to improve the performance of information retrieval systems in settings in which there exists no data and proprietary LLMs may not be available. We hope this can enable the deployment of our system in similar settings. 

%% file: sections/awknowledgements.tex
\section*{Acknowledgements}

We sincerely thank Charlie Dagli, John Holodnak, and Daniel Gwon for their discussion and help in this project. Research was sponsored by the Department of the Air Force Artificial Intelligence Accelerator and was accomplished under Cooperative Agreement Number FA8750-19-2-1000. The views and conclusions contained in this document are those of the authors and should not be interpreted as representing the official policies, either expressed or implied, of the Department of the Air Force or the U.S. Government. The U.S. Government is authorized to reproduce and distribute reprints for Government purposes notwithstanding any copyright notation herein.

%% file: sections/appendix.tex
\begin{table*}[ht!]
   \centering
  \resizebox{\textwidth}{!}{
  \begin{tabular}{l|cc|ccccccc}
    \hline
    \textbf{Model} & DL19 & DL20 & News & Covid & FiQA &  SciFact & DBPedia & NFCorpus &  Robust04 \\
    \hline
    HyDE (Mistral-7B-Instruct)  & 57.8 &	53.9  &	44.0 &	56.9 &	21.6 &	65.1 &	35.3 &	27.7 &	41.5 \\
    & \quad $\pm 0.3$  &	\quad $\pm{0.7}$\ &	\quad $\pm{0.6}$\ &	\quad $\pm{0.8}$\ &	\quad $\pm{0.2}$\ &	\quad $\pm{0.2}$\ &	\quad $\pm{0.3}$\ &	\quad $\pm{0.2}$\ &	\quad $\pm{0.4}$ \\
    \text{HyDE}$_\text{PRF}$ (Mistral-7B-Instruct) & 63.5 &	62.0 &	46.9 &	59.1 &	28.3 &	64.5 &	35.2 &	35.0 &	45.6\ \\
    & \quad $\pm {1.2}$  &	\quad $\pm{0.2}$\ &	\quad $\pm{1.1}$\ &	\quad $\pm{0.5}$\ &	\quad $\pm{0.2}$\ &	\quad $\pm{0.5}$\ &	\quad $\pm{0.1}$\ &	\quad $\pm{0.2}$\ &	\quad $\pm{0.2}$ \\
    ReDE-RF (Default: \text{HyDE}$_\text{PRF}$) & 
    62.8 &	60.4 &	47.1 &	65.6 &	29.3 &	 66.9 &	37.6 &	35.5 &	51.7 \\
    & \quad $\pm 0.0$  &	\quad $\pm{0.1}$\ &	\quad $\pm{0.3}$\ &	\quad $\pm{0.1}$\ &	\quad $\pm{0.1}$\ &	\quad $\pm{0.3}$\ &	\quad $\pm{0.1}$\ &	\quad $\pm{0.1}$\ &	\quad $\pm{0.1}$ \\
  \end{tabular}}
  \caption{NDCG@10 of HyDE-Mistral-7B-Instruct and ReDE-RF (w/ HyDE) with standard deviations across three runs.}
  \label{tab:standard_deviation_across_runs}
\end{table*}

\begin{table*}[ht!]
   \centering
  \resizebox{\textwidth}{!}{
  \begin{tabular}{l|c|cc|ccccccc}
    \hline
    \textbf{Model} & Init Retrieval & DL19 & DL20 & News & Covid & FiQA &  SciFact & DBPedia & NFCorpus &  Robust04 \\
    \hline
    \text{HyDE}$_\text{PRF}$ (Mistral-7B-Instruct) & Hybrid (20) & 63.5 &	62.0 &	46.9 &	59.1 &	28.3 &	64.5 &	35.2 &	35.0 &	45.6 \\
    \text{HyDE}$_\text{PRF}$ (Mistral-7B-Instruct) & Hybrid (10) & 63.6 &	60.5 &	47.2 &	58.8 &	28.3 &	63.4 &	34.6 &	34.6 &	44.5
    
  \end{tabular}}
  \caption{NDCG@10 of \text{HyDE}$_\text{PRF}$ across different number of initially retrieved documents used as context.}
  \label{tab:hyde_prf_diff_init_retrieval}
\end{table*}

\section{Dataset Details}

We show the number of test queries for each dataset used to evaluate ReDE-RF in Table \ref{tab:dataset_details}. 

\begin{table}[H]
  \centering
  \begin{tabular}{l|c}
    \hline
    Dataset & \# Queries \\
    \hline
    TREC DL19 & 43 \\
    TREC DL20 & 54 \\
    TREC-News & 57 \\
    TREC-Covid & 50 \\
    FiQA &  648 \\
    SciFact &  300 \\
    DBPedia & 400 \\
    NFCorpus & 323 \\
    Robust04 & 249 \\
    \hline
  \end{tabular} 
  \caption{Dataset Details}
  \label{tab:dataset_details}
\end{table}

The above datasets have the following licenses. 
\begin{itemize}
    \item TREC DL19 and DL20 are under “MIT License” for non-commercial research purposes.
    \item TREC News and Robust04 are under Copyright.
    \item DBPedia is under CC BY-SA 3.0 license.
    \item SciFact is under CC BY-NC 2.0 license.
    \item TREC Covid is provided under Dataset License Agreement
    \item NFCorpus and FiQA do not report the dataset license as per \citet{thakur2021beir}.
\end{itemize}
The BEIR dataset is under the Apache 2.0 License. 

\section{Model Details}

\begin{itemize}
    \item $\texttt{Mistral-7B-Instruct-v.02}$: A 7B parameter model that is  instruction fine-tuned. Huggingface ID: $\texttt{mistralai/Mistral-7B-Instruct-v0.2}$ 
    \item  $\texttt{contriever}$: Based on $\texttt{bert-base-uncased}$ which has 110M parameters. HuggingFace ID: $\texttt{facebook/contriever}$. 
     \item $\texttt{Mixtral-8x7B-Instruct-v0.1}$: A 8x7B parameter model (47B total parameters) that is instruction fine-tuned. Huggingface ID: $\texttt{mistralai/Mixtral-8x7B-Instruct-v0.1}$ 
    \item $\texttt{gemma-2-2b-it}$: A 2B parameter model that is  instruction fine-tuned. Huggingface ID: $\texttt{google/gemma-2-2b-it}$
    \item $\texttt{gemma-2-9b-it}$: A 9B parameter model that is  instruction fine-tuned. Huggingface ID: $\texttt{google/gemma-2-9b-it}$
    \item $\texttt{Llama-3.2-3B-Instruct}$: A 3B parameter model that is  instruction fine-tuned. Huggingface ID: $\texttt{meta-llama/Llama-3.2-3B-Instruct}$
    \item $\texttt{Llama-3.1-8B-Instruct}$: A 8B parameter model that is  instruction fine-tuned. Huggingface ID: $\texttt{meta-llama/Llama-3.1-8B-Instruct}$ 
    
\end{itemize}

The above models have the following licenses. 
\begin{itemize}
    \item $\texttt{Mistral-7B-Instruct-v0.2}$ is under the Apache 2.0 License. 
    \item $\texttt{Mixtral-8x7B-Instruct-v0.1}$ is under the Apache 2.0 License. 
    \item $\texttt{contriever}$ is under CC BY-NC 4.0 license.
    \item $\texttt{gemma-2-2b-it}$ is under the Apache 2.0 License.
    \item $\texttt{gemma-2-9b-it}$ is under the Apache 2.0 License.
    \item $\texttt{Llama-3.2-3B-Instruct}$ is under the Llama 2 Community License Agreement
    \item $\texttt{Llama-3.1-8B-Instruct}$ is under the Llama 2 Community License Agreement
\end{itemize}
We also leverage Pyserini \cite{lin2021pyserini} which is under the Apache 2.0 License.

\section{Results Across Multiple Runs}
\label{sec:standard_deviations}
Due to the randomness of sampling hypothetical documents from an LLM, we run HyDE, \text{HyDE}$_\text{PRF}$, and ReDE-RF (Default: \text{HyDE}$_\text{PRF}$) three times. In Table \ref{tab:standard_deviation_across_runs} we report the mean and standard deviation of NDCG@10 across the runs for all TREC and BEIR datasets.

\section{HyDE and \text{HyDE}$_\text{PRF}$ Implementation}
\label{sec:hyde_implementation}

\subsection{Generation Details}

To re-implement HyDE and \text{HyDE}$_\text{PRF}$ with Mistral-7B-Instruct we follow the same parameters that were mentioned in the original paper \cite{gao-etal-2023-precise} and in the provided  \href{https://github.com/texttron/hyde/blob/main/src/hyde/generator.py}{codebase}. In particular, we sample eight hypothetical documents from Mistral-7B-Instruct with temperature of 0.7 and allow up to 512 maximum new generation tokens per hypothetical document.

\subsection{\text{HyDE}$_\text{PRF}$ with less in-context documents}
\label{sec:hyde_prf_less_docs}

We explore how less initially retrieved results impact performance of \text{HyDE}$_\text{PRF}$. For results, see Table \ref{tab:hyde_prf_diff_init_retrieval}.

\subsection{Prompts}

For HyDE \cite{gao-etal-2023-precise}, we leverage the same prompts from the original implementation. For \text{HyDE}$_\text{PRF}$, we follow the format of the HyDE prompts and provide context following the format in Q2D/PRF from \citet{jagerman2023query}.

\subsubsection{HyDE Prompts}

\smallskip
\noindent
\textbf{TREC DL19 and DL20}
\begin{tcolorbox}
Please write a passage to answer the question. \\
Question: \{\} \\
Passage: 
\end{tcolorbox}

\noindent
\textbf{SciFact}
\begin{tcolorbox}
Please write a scientific paper passage to support/refute the claim. \\
Claim: \{\} \\
Passage: 
\end{tcolorbox}

\noindent
\textbf{TREC Covid and NFCorpus}
\begin{tcolorbox}
Please write a scientific paper passage to answer the question. \\
Question: \{\} \\
Passage: 
\end{tcolorbox}

\noindent
\textbf{FIQA}
\begin{tcolorbox}
Please write a financial article passage to answer the question. \\
Question: \{\} \\
Passage: 
\end{tcolorbox}

\noindent
\textbf{DBPedia}
\begin{tcolorbox}
Please write a passage to answer the question. \\
Question: \{\} \\
Passage: 
\end{tcolorbox}

\noindent
\textbf{TREC News and Robust04}
\begin{tcolorbox}
Please write a news passage about the topic. \\
Topic: \{\} \\
Passage:
\end{tcolorbox}

\subsubsection{\text{HyDE}$_\text{PRF}$ Prompts}

\smallskip
\noindent
\textbf{TREC DL19 and DL20}
\begin{tcolorbox}
Please write a passage to answer the question based on the context:

Context: \\ 
\{\} \\ 
Question: \{\} \\
Passage: 
\end{tcolorbox}

\noindent
\textbf{SciFact}
\begin{tcolorbox}
Please write a scientific paper passage to support/refute the claim based on the context: 

Context:\\ 
\{\} \\ 
Claim: \{\} \\
Passage: 
\end{tcolorbox}

\noindent
\textbf{TREC Covid and NFCorpus}
\begin{tcolorbox}
Please write a scientific paper passage to answer the question based on the context:

Context:\\ 
\{\} \\ 
Question: \{\} \\
Passage: 
\end{tcolorbox}

\noindent
\textbf{FIQA}
\begin{tcolorbox}
Please write a financial article passage to answer the question based on the context:

Context:\\ 
\{\} \\ 
Question: \{\} \\
Passage: 
\end{tcolorbox}

\noindent
\textbf{DBPedia}
\begin{tcolorbox}
Please write a passage to answer the question based on the context:

Context:\\ 
\{\} \\ 
Question: \{\} \\
Passage: 
\end{tcolorbox}

\noindent
\textbf{TREC News and Robust04}
\begin{tcolorbox}
Please write a news passage about the topic based on the context:

Context:\\ 
\{\} \\ 
Topic: \{\} \\
Passage:
\end{tcolorbox}

\section{Hybrid Retrieval Implementation Details}
\label{sec: hybrid_retrieval_desc}

We use the hybrid retrieval implementation from Pyserini, which scores the document by a weighted average of the sparse retrieval and dense retrieval scores. We implement using the default parameters. See \href{https://github.com/castorini/pyserini/blob/master/pyserini/search/hybrid/_searcher.py}{here} for more details.

\section{DistillReDE Training Details}

\label{sec: distill_rede_training}
To train DistillReDE, we first need synthetic queries given our corpus. We leverage the filtered set of 10K synthetic queries provided by \citet{jeronymo2023inpars} which were generated using GPT-J \cite{wang2021gpt}. Then, we run ReDE-RF on these synthetic queries and store the ReDE-RF embedding for each query. If ReDE-RF finds no relevant documents for a given query, we remove that query from our training set.  

To train DistillReDE, following \citet{pimpalkhute2024softqe}, we optimize the following objective:

\begin{equation}
\mathcal{L}_{\text{DistillReDE}} = 0.5 \mathcal{L}_{\text{MSE}}(\hat{v}_{q_{\textnormal{ReDE}}}, f(q)) + 
    0.5\mathcal{L}_{\text{Cont}}
\end{equation}

where $\mathcal{L}_{\text{Cont}}$ is a contrastive objective \cite{karpukhin-etal-2020-dense}:

\begin{eqnarray}
&& \scalebox{1}{$
    L(q_i, \hat{v}_{q_{\textnormal{ReDE}, {i}}}^+,
    \hat{v}_{q_{\textnormal{ReDE}, {i,1}}}^-, \dots, 
    \hat{v}_{q_{\textnormal{ReDE}, {i,n}}}^-)
    $} \\
&=& \scalebox{1}{$
    -\log \frac{ e^{\mathrm{sim}(q_i, \hat{v}_{q_{\textnormal{ReDE}, {i}}}^+)} }
    { e^{\mathrm{sim}(q_i,  \hat{v}_{q_{\textnormal{ReDE}, {i}}}^+)} 
    + \sum\limits_{j=1}^n{e^{\mathrm{sim}(q_i, 
    \hat{v}_{q_{\textnormal{ReDE}, {i,j}}}^-)}} }
    $} \nonumber
\end{eqnarray}

where given a query, $q$, we have one positive ReDE-RF embedding $\hat{v}_{q_{\textnormal{ReDE}, {i}}}^+$ and $n$ negative ReDE-RF embeddings $(\hat{v}_{q_{\textnormal{ReDE}, {i,1}}}^-, \cdots, \hat{v}_{q_{\textnormal{ReDE}, {i,n}}}^-)$.

We train with in-batch negatives using a training batch size of 256 and use the Adam optimizer \cite{diederik2014adam} with a learning rate of 5e-5.

\section{ReDE-RF Prompt Varations}
\label{sec:rede_prompt_variations}
Below are the prompts studied in the prompt variations subsection of Section \ref{sec:Initial_Retrieval_Abalation}.

\noindent
\textbf{pointwise.yes\_no \cite{zhuang2024setwise}}

\begin{tcolorbox}
Passage: \{\} \\ 
Query: \{\} \\ 
Does the passage answer the query? Answer 'Yes' or 'No'.
\end{tcolorbox}

\noindent
\textbf{ RG-YN \cite{zhuang2024beyond}}

\begin{tcolorbox}
For the following query and document, judge whether they are relevant. Output “Yes” or “No”. \\ 
\\ 
Query: \{\} \\ 
Document: \{\} \\ 
Output:
\end{tcolorbox}

\noindent
\textbf{RG-YN$*$}

\begin{tcolorbox}
For the following query and document, judge whether they are relevant. Output “Yes” if the passage is dedicated to the query and contains the exact answer 
and output "No" if the passage has nothing to do with the query. \\
\\
Query: \{\} \\ 
Document: \{\} \\ 
Output:
\end{tcolorbox}

\noindent
\textbf{\citet{thomas2024large}}

\begin{tcolorbox}
You are a search quality rater evaluating the relevance of web pages. Given a query and a web page, you must provide a score on an integer scale of 0 to 1 with the following meanings:
\\
\\
1 = highly relevant, very helpful for this query \\
0 = not relevant, should never be shown for this query
\\
\\
Assume that you are writing a report on the subject of the topic. If the web page is primarily about the topic, or contains vital information about the topic, mark it 1. Otherwise, mark it 0.
\\

Passage: \{\} \\ 
Query: \{\} \\ 
Score:
\end{tcolorbox}

%% file: main.bbl
\begin{thebibliography}{44}
\providecommand{\natexlab}[1]{#1}

\bibitem[{Chuang et~al.(2023)Chuang, Fang, Li, Yih, and Glass}]{chuang-etal-2023-expand}
Yung-Sung Chuang, Wei Fang, Shang-Wen Li, Wen-tau Yih, and James Glass. 2023.
\newblock \href {https://doi.org/10.18653/v1/2023.findings-acl.768} {Expand, rerank, and retrieve: Query reranking for open-domain question answering}.
\newblock In \emph{Findings of the Association for Computational Linguistics: ACL 2023}, pages 12131--12147, Toronto, Canada. Association for Computational Linguistics.

\bibitem[{Craswell et~al.(2021)Craswell, Mitra, Yilmaz, and Campos}]{craswell2021overview}
Nick Craswell, Bhaskar Mitra, Emine Yilmaz, and Daniel Campos. 2021.
\newblock \href {https://arxiv.org/abs/2102.07662} {Overview of the trec 2020 deep learning track}.
\newblock \emph{Preprint}, arXiv:2102.07662.

\bibitem[{Craswell et~al.(2020)Craswell, Mitra, Yilmaz, Campos, and Voorhees}]{craswell2020overview}
Nick Craswell, Bhaskar Mitra, Emine Yilmaz, Daniel Campos, and Ellen~M. Voorhees. 2020.
\newblock \href {https://arxiv.org/abs/2003.07820} {Overview of the trec 2019 deep learning track}.
\newblock \emph{Preprint}, arXiv:2003.07820.

\bibitem[{Dai et~al.(2022)Dai, Zhao, Ma, Luan, Ni, Lu, Bakalov, Guu, Hall, and Chang}]{dai2022promptagator}
Zhuyun Dai, Vincent~Y Zhao, Ji~Ma, Yi~Luan, Jianmo Ni, Jing Lu, Anton Bakalov, Kelvin Guu, Keith~B Hall, and Ming-Wei Chang. 2022.
\newblock Promptagator: Few-shot dense retrieval from 8 examples.
\newblock \emph{arXiv preprint arXiv:2209.11755}.

\bibitem[{Dao et~al.(2022)Dao, Fu, Ermon, Rudra, and R{\'e}}]{dao2022flashattention}
Tri Dao, Dan Fu, Stefano Ermon, Atri Rudra, and Christopher R{\'e}. 2022.
\newblock Flashattention: Fast and memory-efficient exact attention with io-awareness.
\newblock \emph{Advances in Neural Information Processing Systems}, 35:16344--16359.

\bibitem[{Devlin et~al.(2019)Devlin, Chang, Lee, and Toutanova}]{devlin-etal-2019-bert}
Jacob Devlin, Ming-Wei Chang, Kenton Lee, and Kristina Toutanova. 2019.
\newblock \href {https://doi.org/10.18653/v1/N19-1423} {{BERT}: Pre-training of deep bidirectional transformers for language understanding}.
\newblock In \emph{Proceedings of the 2019 Conference of the North {A}merican Chapter of the Association for Computational Linguistics: Human Language Technologies, Volume 1 (Long and Short Papers)}, pages 4171--4186, Minneapolis, Minnesota. Association for Computational Linguistics.

\bibitem[{Diederik(2014)}]{diederik2014adam}
P~Kingma Diederik. 2014.
\newblock Adam: A method for stochastic optimization.
\newblock \emph{(No Title)}.

\bibitem[{Gao et~al.(2023)Gao, Ma, Lin, and Callan}]{gao-etal-2023-precise}
Luyu Gao, Xueguang Ma, Jimmy Lin, and Jamie Callan. 2023.
\newblock \href {https://doi.org/10.18653/v1/2023.acl-long.99} {Precise zero-shot dense retrieval without relevance labels}.
\newblock In \emph{Proceedings of the 61st Annual Meeting of the Association for Computational Linguistics (Volume 1: Long Papers)}, pages 1762--1777, Toronto, Canada. Association for Computational Linguistics.

\bibitem[{Izacard et~al.(2021)Izacard, Caron, Hosseini, Riedel, Bojanowski, Joulin, and Grave}]{izacard2021unsupervised}
Gautier Izacard, Mathilde Caron, Lucas Hosseini, Sebastian Riedel, Piotr Bojanowski, Armand Joulin, and Edouard Grave. 2021.
\newblock Unsupervised dense information retrieval with contrastive learning.
\newblock \emph{arXiv preprint arXiv:2112.09118}.

\bibitem[{Jagerman et~al.(2023)Jagerman, Zhuang, Qin, Wang, and Bendersky}]{jagerman2023query}
Rolf Jagerman, Honglei Zhuang, Zhen Qin, Xuanhui Wang, and Michael Bendersky. 2023.
\newblock Query expansion by prompting large language models.
\newblock \emph{arXiv preprint arXiv:2305.03653}.

\bibitem[{Jeronymo et~al.(2023)Jeronymo, Bonifacio, Abonizio, Fadaee, Lotufo, Zavrel, and Nogueira}]{jeronymo2023inpars}
Vitor Jeronymo, Luiz Bonifacio, Hugo Abonizio, Marzieh Fadaee, Roberto Lotufo, Jakub Zavrel, and Rodrigo Nogueira. 2023.
\newblock Inpars-v2: Large language models as efficient dataset generators for information retrieval.
\newblock \emph{arXiv preprint arXiv:2301.01820}.

\bibitem[{Jiang et~al.(2023)Jiang, Sablayrolles, Mensch, Bamford, Chaplot, Casas, Bressand, Lengyel, Lample, Saulnier et~al.}]{jiang2023mistral}
Albert~Q Jiang, Alexandre Sablayrolles, Arthur Mensch, Chris Bamford, Devendra~Singh Chaplot, Diego de~las Casas, Florian Bressand, Gianna Lengyel, Guillaume Lample, Lucile Saulnier, et~al. 2023.
\newblock Mistral 7b.
\newblock \emph{arXiv preprint arXiv:2310.06825}.

\bibitem[{Karpukhin et~al.(2020)Karpukhin, Oguz, Min, Lewis, Wu, Edunov, Chen, and Yih}]{karpukhin-etal-2020-dense}
Vladimir Karpukhin, Barlas Oguz, Sewon Min, Patrick Lewis, Ledell Wu, Sergey Edunov, Danqi Chen, and Wen-tau Yih. 2020.
\newblock \href {https://doi.org/10.18653/v1/2020.emnlp-main.550} {Dense passage retrieval for open-domain question answering}.
\newblock In \emph{Proceedings of the 2020 Conference on Empirical Methods in Natural Language Processing (EMNLP)}, pages 6769--6781, Online. Association for Computational Linguistics.

\bibitem[{Lee et~al.(2024)Lee, Dai, Ren, Chen, Cer, Cole, Hui, Boratko, Kapadia, Ding et~al.}]{lee2024gecko}
Jinhyuk Lee, Zhuyun Dai, Xiaoqi Ren, Blair Chen, Daniel Cer, Jeremy~R Cole, Kai Hui, Michael Boratko, Rajvi Kapadia, Wen Ding, et~al. 2024.
\newblock Gecko: Versatile text embeddings distilled from large language models.
\newblock \emph{arXiv preprint arXiv:2403.20327}.

\bibitem[{Lei et~al.(2024)Lei, Cao, Zhou, Shen, and Yates}]{lei-etal-2024-corpus}
Yibin Lei, Yu~Cao, Tianyi Zhou, Tao Shen, and Andrew Yates. 2024.
\newblock \href {https://aclanthology.org/2024.eacl-short.34} {Corpus-steered query expansion with large language models}.
\newblock In \emph{Proceedings of the 18th Conference of the European Chapter of the Association for Computational Linguistics (Volume 2: Short Papers)}, pages 393--401, St. Julian{'}s, Malta. Association for Computational Linguistics.

\bibitem[{Li et~al.(2022)Li, Zhuang, Ma, Lin, and Zuccon}]{li2022pseudo}
Hang Li, Shengyao Zhuang, Xueguang Ma, Jimmy Lin, and Guido Zuccon. 2022.
\newblock Pseudo-relevance feedback with dense retrievers in pyserini.
\newblock In \emph{Proceedings of the 26th Australasian Document Computing Symposium}, pages 1--6.

\bibitem[{Lin et~al.(2021)Lin, Ma, Lin, Yang, Pradeep, and Nogueira}]{lin2021pyserini}
Jimmy Lin, Xueguang Ma, Sheng-Chieh Lin, Jheng-Hong Yang, Ronak Pradeep, and Rodrigo Nogueira. 2021.
\newblock Pyserini: A python toolkit for reproducible information retrieval research with sparse and dense representations.
\newblock In \emph{Proceedings of the 44th International ACM SIGIR Conference on Research and Development in Information Retrieval}, pages 2356--2362.

\bibitem[{Lin et~al.(2022)Lin, Nogueira, and Yates}]{lin2022pretrained}
Jimmy Lin, Rodrigo Nogueira, and Andrew Yates. 2022.
\newblock \emph{Pretrained transformers for text ranking: Bert and beyond}.
\newblock Springer Nature.

\bibitem[{Liu et~al.(2024)Liu, Lin, Hewitt, Paranjape, Bevilacqua, Petroni, and Liang}]{liu2024lost}
Nelson~F Liu, Kevin Lin, John Hewitt, Ashwin Paranjape, Michele Bevilacqua, Fabio Petroni, and Percy Liang. 2024.
\newblock Lost in the middle: How language models use long contexts.
\newblock \emph{Transactions of the Association for Computational Linguistics}, 12:157--173.

\bibitem[{Mackie et~al.(2023)Mackie, Chatterjee, and Dalton}]{mackie2023generative}
Iain Mackie, Shubham Chatterjee, and Jeffrey Dalton. 2023.
\newblock Generative relevance feedback with large language models.
\newblock In \emph{Proceedings of the 46th International ACM SIGIR Conference on Research and Development in Information Retrieval}, pages 2026--2031.

\bibitem[{Mao et~al.(2021)Mao, He, Liu, Shen, Gao, Han, and Chen}]{mao-etal-2021-generation}
Yuning Mao, Pengcheng He, Xiaodong Liu, Yelong Shen, Jianfeng Gao, Jiawei Han, and Weizhu Chen. 2021.
\newblock \href {https://doi.org/10.18653/v1/2021.acl-long.316} {Generation-augmented retrieval for open-domain question answering}.
\newblock In \emph{Proceedings of the 59th Annual Meeting of the Association for Computational Linguistics and the 11th International Joint Conference on Natural Language Processing (Volume 1: Long Papers)}, pages 4089--4100, Online. Association for Computational Linguistics.

\bibitem[{Nogueira et~al.(2020)Nogueira, Jiang, Pradeep, and Lin}]{nogueira-etal-2020-document}
Rodrigo Nogueira, Zhiying Jiang, Ronak Pradeep, and Jimmy Lin. 2020.
\newblock \href {https://doi.org/10.18653/v1/2020.findings-emnlp.63} {Document ranking with a pretrained sequence-to-sequence model}.
\newblock In \emph{Findings of the Association for Computational Linguistics: EMNLP 2020}, pages 708--718, Online. Association for Computational Linguistics.

\bibitem[{Pimpalkhute et~al.(2024)Pimpalkhute, Heyer, Yin, and Gupta}]{pimpalkhute2024softqe}
Varad Pimpalkhute, John Heyer, Xusen Yin, and Sameer Gupta. 2024.
\newblock Softqe: Learned representations of queries expanded by llms.
\newblock In \emph{European Conference on Information Retrieval}, pages 68--77. Springer.

\bibitem[{Qu et~al.(2020)Qu, Ding, Liu, Liu, Ren, Zhao, Dong, Wu, and Wang}]{qu2020rocketqa}
Yingqi Qu, Yuchen Ding, Jing Liu, Kai Liu, Ruiyang Ren, Wayne~Xin Zhao, Daxiang Dong, Hua Wu, and Haifeng Wang. 2020.
\newblock Rocketqa: An optimized training approach to dense passage retrieval for open-domain question answering.
\newblock \emph{arXiv preprint arXiv:2010.08191}.

\bibitem[{Robertson et~al.(2009)Robertson, Zaragoza et~al.}]{robertson2009probabilistic}
Stephen Robertson, Hugo Zaragoza, et~al. 2009.
\newblock The probabilistic relevance framework: Bm25 and beyond.
\newblock \emph{Foundations and Trends{\textregistered} in Information Retrieval}, 3(4):333--389.

\bibitem[{Sachan et~al.(2023)Sachan, Lewis, Yogatama, Zettlemoyer, Pineau, and Zaheer}]{sachan2023questions}
Devendra~Singh Sachan, Mike Lewis, Dani Yogatama, Luke Zettlemoyer, Joelle Pineau, and Manzil Zaheer. 2023.
\newblock Questions are all you need to train a dense passage retriever.
\newblock \emph{Transactions of the Association for Computational Linguistics}, 11:600--616.

\bibitem[{Shen et~al.(2024)Shen, Long, Geng, Tao, Lei, Zhou, Blumenstein, and Jiang}]{shen2024retrieval}
Tao Shen, Guodong Long, Xiubo Geng, Chongyang Tao, Yibin Lei, Tianyi Zhou, Michael Blumenstein, and Daxin Jiang. 2024.
\newblock Retrieval-augmented retrieval: Large language models are strong zero-shot retriever.
\newblock In \emph{Findings of the Association for Computational Linguistics ACL 2024}, pages 15933--15946.

\bibitem[{Shi et~al.(2023)Shi, Han, Lewis, Tsvetkov, Zettlemoyer, and Yih}]{shi2023trusting}
Weijia Shi, Xiaochuang Han, Mike Lewis, Yulia Tsvetkov, Luke Zettlemoyer, and Scott Wen-tau Yih. 2023.
\newblock Trusting your evidence: Hallucinate less with context-aware decoding.
\newblock \emph{arXiv preprint arXiv:2305.14739}.

\bibitem[{Simhi et~al.(2024)Simhi, Herzig, Szpektor, and Belinkov}]{simhi2024constructing}
Adi Simhi, Jonathan Herzig, Idan Szpektor, and Yonatan Belinkov. 2024.
\newblock Constructing benchmarks and interventions for combating hallucinations in llms.
\newblock \emph{arXiv preprint arXiv:2404.09971}.

\bibitem[{Springer et~al.(2024)Springer, Kotha, Fried, Neubig, and Raghunathan}]{springer2024repetition}
Jacob~Mitchell Springer, Suhas Kotha, Daniel Fried, Graham Neubig, and Aditi Raghunathan. 2024.
\newblock Repetition improves language model embeddings.
\newblock \emph{arXiv preprint arXiv:2402.15449}.

\bibitem[{Thakur et~al.(2021)Thakur, Reimers, R{\"u}ckl{\'e}, Srivastava, and Gurevych}]{thakur2021beir}
Nandan Thakur, Nils Reimers, Andreas R{\"u}ckl{\'e}, Abhishek Srivastava, and Iryna Gurevych. 2021.
\newblock \href {https://openreview.net/forum?id=wCu6T5xFjeJ} {{BEIR}: A heterogeneous benchmark for zero-shot evaluation of information retrieval models}.
\newblock In \emph{Thirty-fifth Conference on Neural Information Processing Systems Datasets and Benchmarks Track (Round 2)}.

\bibitem[{Thomas et~al.(2024)Thomas, Spielman, Craswell, and Mitra}]{thomas2024large}
Paul Thomas, Seth Spielman, Nick Craswell, and Bhaskar Mitra. 2024.
\newblock Large language models can accurately predict searcher preferences.
\newblock In \emph{Proceedings of the 47th International ACM SIGIR Conference on Research and Development in Information Retrieval}, pages 1930--1940.

\bibitem[{Upadhyay et~al.(2024)Upadhyay, Kamalloo, and Lin}]{upadhyay2024llms}
Shivani Upadhyay, Ehsan Kamalloo, and Jimmy Lin. 2024.
\newblock Llms can patch up missing relevance judgments in evaluation.
\newblock \emph{arXiv preprint arXiv:2405.04727}.

\bibitem[{Wang and Komatsuzaki(2021)}]{wang2021gpt}
Ben Wang and Aran Komatsuzaki. 2021.
\newblock Gpt-j-6b: A 6 billion parameter autoregressive language model.

\bibitem[{Wang et~al.(2023{\natexlab{a}})Wang, Yang, Huang, Yang, Majumder, and Wei}]{wang2023improving}
Liang Wang, Nan Yang, Xiaolong Huang, Linjun Yang, Rangan Majumder, and Furu Wei. 2023{\natexlab{a}}.
\newblock Improving text embeddings with large language models.
\newblock \emph{arXiv preprint arXiv:2401.00368}.

\bibitem[{Wang et~al.(2023{\natexlab{b}})Wang, Yang, and Wei}]{wang-etal-2023-query2doc}
Liang Wang, Nan Yang, and Furu Wei. 2023{\natexlab{b}}.
\newblock \href {https://doi.org/10.18653/v1/2023.emnlp-main.585} {Query2doc: Query expansion with large language models}.
\newblock In \emph{Proceedings of the 2023 Conference on Empirical Methods in Natural Language Processing}, pages 9414--9423, Singapore. Association for Computational Linguistics.

\bibitem[{Wolf et~al.(2019)Wolf, Debut, Sanh, Chaumond, Delangue, Moi, Cistac, Rault, Louf, Funtowicz et~al.}]{wolf2019huggingface}
Thomas Wolf, Lysandre Debut, Victor Sanh, Julien Chaumond, Clement Delangue, Anthony Moi, Pierric Cistac, Tim Rault, R{\'e}mi Louf, Morgan Funtowicz, et~al. 2019.
\newblock Huggingface's transformers: State-of-the-art natural language processing.
\newblock \emph{arXiv preprint arXiv:1910.03771}.

\bibitem[{Xiong et~al.(2020)Xiong, Xiong, Li, Tang, Liu, Bennett, Ahmed, and Overwijk}]{xiong2020approximate}
Lee Xiong, Chenyan Xiong, Ye~Li, Kwok-Fung Tang, Jialin Liu, Paul Bennett, Junaid Ahmed, and Arnold Overwijk. 2020.
\newblock Approximate nearest neighbor negative contrastive learning for dense text retrieval.
\newblock \emph{arXiv preprint arXiv:2007.00808}.

\bibitem[{Xiong et~al.(2021)Xiong, Xiong, Li, Tang, Liu, Bennett, Ahmed, and Overwijk}]{xiong2021approximate}
Lee Xiong, Chenyan Xiong, Ye~Li, Kwok-Fung Tang, Jialin Liu, Paul~N. Bennett, Junaid Ahmed, and Arnold Overwijk. 2021.
\newblock \href {https://openreview.net/forum?id=zeFrfgyZln} {Approximate nearest neighbor negative contrastive learning for dense text retrieval}.
\newblock In \emph{International Conference on Learning Representations}.

\bibitem[{Yoon et~al.(2024)Yoon, Kim, Jeon, Kim, Jo, and Kang}]{yoon2024ask}
Chanwoong Yoon, Gangwoo Kim, Byeongguk Jeon, Sungdong Kim, Yohan Jo, and Jaewoo Kang. 2024.
\newblock Ask optimal questions: Aligning large language models with retriever's preference in conversational search.
\newblock \emph{arXiv preprint arXiv:2402.11827}.

\bibitem[{Zhou et~al.(2023)Zhou, Zhang, Poon, and Chen}]{zhou-etal-2023-context}
Wenxuan Zhou, Sheng Zhang, Hoifung Poon, and Muhao Chen. 2023.
\newblock \href {https://doi.org/10.18653/v1/2023.findings-emnlp.968} {Context-faithful prompting for large language models}.
\newblock In \emph{Findings of the Association for Computational Linguistics: EMNLP 2023}, pages 14544--14556, Singapore. Association for Computational Linguistics.

\bibitem[{Zhuang et~al.(2024{\natexlab{a}})Zhuang, Qin, Hui, Wu, Yan, Wang, and Bendersky}]{zhuang2024beyond}
Honglei Zhuang, Zhen Qin, Kai Hui, Junru Wu, Le~Yan, Xuanhui Wang, and Michael Bendersky. 2024{\natexlab{a}}.
\newblock Beyond yes and no: Improving zero-shot llm rankers via scoring fine-grained relevance labels.
\newblock In \emph{Proceedings of the 2024 Conference of the North American Chapter of the Association for Computational Linguistics: Human Language Technologies (Volume 2: Short Papers)}, pages 358--370.

\bibitem[{Zhuang et~al.(2024{\natexlab{b}})Zhuang, Ma, Koopman, Lin, and Zuccon}]{zhuang2024promptreps}
Shengyao Zhuang, Xueguang Ma, Bevan Koopman, Jimmy Lin, and Guido Zuccon. 2024{\natexlab{b}}.
\newblock Promptreps: Prompting large language models to generate dense and sparse representations for zero-shot document retrieval.
\newblock \emph{arXiv preprint arXiv:2404.18424}.

\bibitem[{Zhuang et~al.(2024{\natexlab{c}})Zhuang, Zhuang, Koopman, and Zuccon}]{zhuang2024setwise}
Shengyao Zhuang, Honglei Zhuang, Bevan Koopman, and Guido Zuccon. 2024{\natexlab{c}}.
\newblock A setwise approach for effective and highly efficient zero-shot ranking with large language models.
\newblock In \emph{Proceedings of the 47th International ACM SIGIR Conference on Research and Development in Information Retrieval}, pages 38--47.

\end{thebibliography}
